\documentclass[12pt,preprint]{aastex}

\slugcomment{Submitted to the The Astrophysical Journal Letters}

\shorttitle{Detecting Moons of Pulsar Planets} \shortauthors{Lewis,
Sackett \& Mardling}

\begin{document}

\title{Possibility of Detecting Moons of Pulsar Planets Through Time-of-Arrival Analysis}

\author{Karen M. Lewis}
\affil{School of Mathematical Sciences, Monash University, Clayton,
Victoria 3800, Australia; karen.lewis@sci.monash.edu.au}
\author{Penny D. Sackett}
\affil{Research School of Astronomy and Astrophysics, Australian
National University, Mt. Stromlo Observatory, Cotter Road, Weston,
ACT 2611, Australia; Penny.Sackett@anu.edu.au}
\and
\author{Rosemary A. Mardling}
\affil{School of Mathematical Sciences, Monash University, Clayton,
Victoria 3800, Australia; rosemary.mardling@sci.monash.edu.au}

\begin{abstract}
The perturbation caused by planet-moon binarity on the
time-of-arrival signal of a pulsar with an orbiting planet is
derived for the case in which the orbits of the moon and the planet-moon
barycenter are both circular and coplanar. The signal consists of
two sinusoids with frequency $(2n_p - 3n_b )$ and $(2n_p - n_b )$,
where $n_p$ and $n_b$ are the mean motions of the planet and
moon around their barycenter, and the planet-moon system around the host, respectively.
The amplitude of the signal is equal to the fraction $\sin I[9(M_p M_m)/16(M_p + M_m)^2] [r/R]^5$
of the system crossing time $R/c$, where $M_p$ and $M_m$ are the the masses of the planet and moon, 
$r$ is their orbital separation, $R$ is the distance between the host pulsar and planet-moon barycenter, $I$ is the inclination of the orbital plane of the planet, and $c$ is the speed of light.   The analysis is
applied to the case of PSR~B1620-26~b, a pulsar planet, to constrain
the orbital separation and mass of any possible moons.  We find that a stable moon orbiting this pulsar planet could be detected, if the moon had a separation of about one fiftieth of that of the orbit of the planet around the pulsar, and a mass ratio to the planet of $\sim$ 5\% or larger.

\end{abstract}


\section{Introduction to Extra-solar Moons}

In the past decade and a half, over two hundred and fifty
extra-solar planets have been discovered\footnote{See, for example,
http://exoplanet.eu/catalogue.php}. With the data expected to be
produced by satellites such as COROT \citep{Auvergneetal2003} and
Kepler \citep*{Basrietal2005}, it will not only be possible to find
smaller planets, but moons of those planets as well
\citep{Szaboetal2006}. As a result, the detectability of extra-solar
moons is starting to be explored in terms of their effect on
planetary microlensing \citep{Hanetal2002} and transit lightcurves
\citep{Sartorettietal1999, Szaboetal2006}. Upper limits
have already been placed on the mass and radius of putative moons of
the planets HD~209458~b \citep{Brownetal2001}, OGLE-TR-113 b \citep{Gillonetal2006} and
HD~189733~b \citep{Pontetal2007}.

While the limitations of microlensing and the transit technique for
detecting moons have been discussed and used in the literature, the
limitations of other techniques such as the time-of-arrival (TOA)
technique have not. This technique involves determining the
variations in line-of-sight position to the host star, usually a
pulsar, using the observed time of periodic events associated with
that host. The aim of this analysis is to explore exactly what the
TOA signal of a planet-moon pair is, and relate it to the planetary
systems that can give the most precise timing information, the
systems around millisecond pulsars.

\section{Review of Planetary Detection Around Millisecond Pulsars}

The first planetary system outside the Solar System was detected
around the millisecond pulsar PSR~1257+12 \citep{Wolszczanetal1992}.
This detection was made by investigating periodic variations in the
time of arrival of its radio pulses using a timing model.  An example timing model for the case in which the planet's orbit around the pulsar is circular is:
\begin{equation}\label{TOA_example}
\left(t_N - t_0\right) = \left(T_N - T_0\right) + \Delta T_{corr}
 + TOA_{pert,p}(M_s, M_p, R, I, \phi_{b}(0)),
\end{equation}
where $t_0$ and $t_N$ are the
times at which the initial and $N^{th}$ pulses are emitted in the pulsar's
frame, $T_0$ and $T_N$ are the times the initial and $N^{th}$ pulses
are received in the observatory's frame, and the term $\Delta
T_{corr}$ acts to change the frame of reference from the
observatory on Earth to the barycenter of the pulsar system \citep[see][for a more complete discussion of the components of $\Delta T_{corr}$]{Backer1993}. The final term represents the effect of a planet on the motion of the pulsar, where $R$ is the planet-pulsar distance, $I$ is the angle between the normal of the planet-pulsar orbit and the line-of-sight and $\phi_b(0)$ is the initial angular position of the planet measured from the $x$-axis, about the system barycenter. In addition, $M_s$ and $M_p$ are the mass of the pulsar and the planet, respectively.

Currently, four planets around two millisecond pulsars have been
discovered, three around PSR~1257+12
\citep{Wolszczanetal1992,Wolszczan1994} and one around PSR~B1620-26
\citep*{Backeretal1993}. These four planets include one with mass
$~0.02$ Earth masses, the lowest mass extra-solar planet known. This
low-mass detection threshold, in addition to measurements of orbital
perturbations \citep[e.g., the 2:3 orbital resonance between
PSR~1257+12c and PSR~1257+12d,][]{Konackietal2003}, demonstrate
the sensitivity of the TOA technique. Indeed, as a result of their
high rotation rate (and thus large number of sampled pulses) and
low level of noise activity, millisecond pulsars make optimal
targets for high precision TOA work \citep[see e.g.,][]{Cordes1993}.

\section{What is the TOA Perturbation Caused by a Moon?}

\begin{figure}[tb]
\begin{center}
\includegraphics[height=2.28in,width=3.84in]{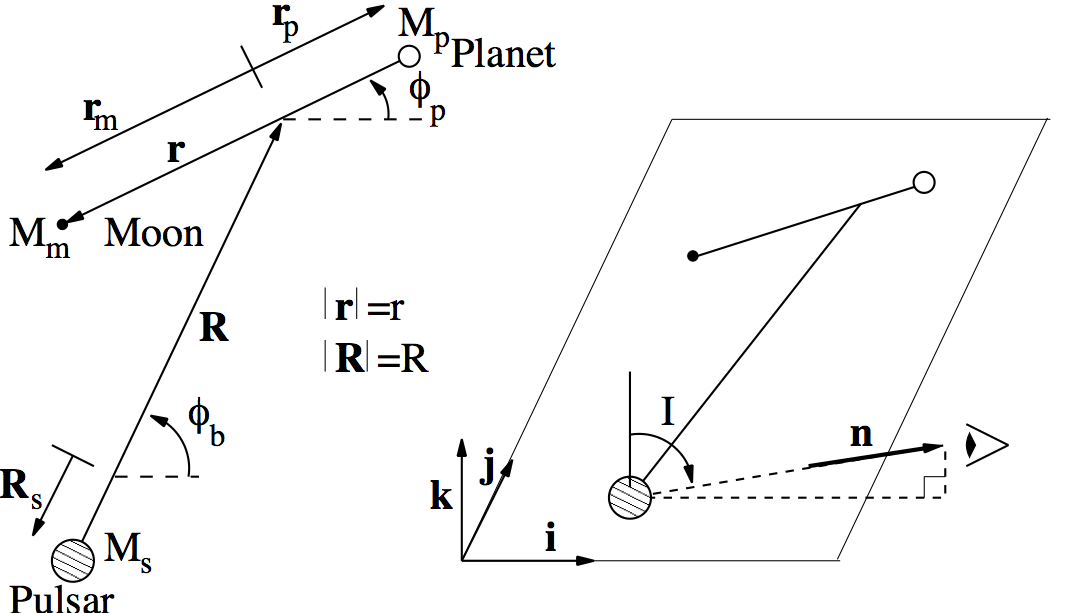}
\caption{Coordinate system used in the analysis of the TOA perturbation
caused by a moon. The subscripts s, b, p and m refer to the pulsar,
planet-moon barycenter, planet and moon, respectively. The diagram on the left shows the quantities used to describe the position of the three bodies in their mutual orbital
plane, while the diagram on the right shows the relationship between
this orbital plane and the observer.}
\label{CoordSystem}
\end{center}
\end{figure}

In order to investigate the perturbation caused by planet-moon binarity, the
timing model presented in equation~(\ref{TOA_example}) must be
updated to include effects due to the presence of the moon. For simplicity, we consider here only systems in which both the orbit of the planet and moon around their common barycenter, and the orbit of the planet-moon barycenter around the pulsar, are both circular and lie in the same plane.
An example model taking this assumption into account is:
\begin{eqnarray}
\left(t_N - t_0\right) = \left(T_N - T_0\right) &+& \Delta T_{corr}
+ TOA_{pert,p}(M_s, M_p+M_m, R, I, \phi_{b}(0)) \nonumber\\
&+&
TOA_{pert,pm}(M_s, M_p, M_m, r, R, I, \phi_{b}(0),\phi_{p}(0)).\label{TOA_example_m}
\end{eqnarray}
We have explicitly modified $TOA_{pert,p}$ to indicate that it
depends on the combined planet-moon mass, and included another term,
$TOA_{pert,pm}$, to account for planet-moon binarity.  Here $M_m$ is
the mass of the moon, $r$ is the distance between the planet and the moon, and $\phi_p(0)$ is the initial angular position of the planet measured from the planet-moon barycenter.  The quantities $R$, $r$, $I$, $\phi_b$ and $\phi_p$ are shown in Figure \ref{CoordSystem}.  The functional form of $TOA_{pert,pm}$ can be derived from $\mathbf{R}_s$, the vector between the system barycenter and the pulsar, using:
\begin{equation}
\frac{1}{c}\int_0^t\int_0^{t'} \mathbf{\ddot{R}}_s \cdot \mathbf{n} dt'' dt' =
TOA_{pert,p} + TOA_{pert,pm},\label{TOA-TOApertpmdef}
\end{equation}
where $c$ is the speed of light and $\mathbf{n}$ is a unit vector pointing along the line of sight, the only direction along which quantities can be measured.

The governing equation for $\mathbf{R_s}$ can be written as the sum of the zeroth order term, which describes $TOA_{pert,p}$, and the tidal terms, which describe $TOA_{pert,pm}$:
\begin{eqnarray}
\frac{d^2\mathbf{R}_s}{dt^2} &=& \frac{G (M_p +
M_m)}{R^3}\mathbf{R} \nonumber\\
&+& \left[- \frac{G (M_p + M_m)}{R^3}\mathbf{R} + \frac{G M_p}{\left|\mathbf{R} 
+ \mathbf{r}_p\right|^3}(\mathbf{R}
+ \mathbf{r}_p) + \frac{G M_m}{\left|\mathbf{R} +
\mathbf{r}_m\right|^3}(\mathbf{R} + \mathbf{r}_m)\right],\label{TOA-d2Rs}
\end{eqnarray}
where the tidal terms have been collected into square brackets and noting that $\mathbf{R}_s = - (M_p + M_m)/(M_s + M_p + M_m)\mathbf{R}$, $\mathbf{r}_p = - M_m/(M_p + M_m)\mathbf{r}$ and $\mathbf{r}_m =
M_p/(M_p + M_m)\mathbf{r}$, and where the vectors $\mathbf{R}_s$, $\mathbf{R}$, $\mathbf{r}_p$, $\mathbf{r}_m$ and $\mathbf{r}$ are also shown in Figure \ref{CoordSystem}. Using
the coordinate system in Figure \ref{CoordSystem}, it can be
seen after some algebra that:
\begin{eqnarray}
\label{TOA-Vec-Rm} \mathbf{R} + \mathbf{r}_m &=& \left[ R\cos
\phi_b - \frac{M_p}{M_p+M_m}r \cos \phi_p\right]\mathbf{i} 
+ \left[R\sin \phi_b - \frac{M_p}{M_p+M_m}r \sin \phi_p
\right]\mathbf{j}\\
\label{TOA-Vec-Rp} \mathbf{R} + \mathbf{r}_p &=& \left[R\cos
\phi_b + \frac{M_m}{M_p+M_m}r \cos \phi_p\right]\mathbf{i} 
+ \left[R \sin \phi_b + \frac{M_m}{M_p+M_m}r \sin \phi_p
\right]\mathbf{j}
\end{eqnarray}
where $\mathbf{i}$ and $\mathbf{j}$ are defined in Figure
\ref{CoordSystem}, and $\mathbf{i}$ is the direction to the
line-of-sight, projected onto the plane of the orbit.

As the orbits are both circular and coplanar, we have that
$\phi_b(t) = n_b t + \phi_b(0)$, and $\phi_p(t) = n_p t + \phi_p(0)$ where
$n_b$ and $n_p$ are the constant mean motions of the two respective orbits.  Substituting
equations~(\ref{TOA-Vec-Rm}) and (\ref{TOA-Vec-Rp}) into the
coefficients of the last two terms of equation~(\ref{TOA-d2Rs}),
assuming $r \ll R$ and using the binomial expansion to
order $r^2/R^2$ gives:
\begin{eqnarray}
 \frac{G M_m}{\left|\mathbf{R} +
\mathbf{r}_m\right|^3} & = & \frac{G M_m}{R^3}\left[1 +
3\frac{M_p}{(M_m + M_p)}\frac{r}{R}\cos(\phi_b -
\phi_p) \right. \nonumber \\ 
 && \left. + \frac{M_p^2}{(M_m +
M_p)^2}\frac{r^2}{R^2} \left(-\frac{3}{2} + \frac{15}{2}
\cos^2(\phi_b - \phi_p)\right)\right]
\label{TOA-expnd-ms}
\end{eqnarray}
\begin{eqnarray}
\frac{G M_p}{\left|\mathbf{R} +
\mathbf{r}_p\right|^3} &=& \frac{G M_p}{R^3}\left[1 -
3\frac{M_m}{(M_m + M_p)}\frac{r}{R}\cos(\phi_b - \phi_p) \right. \nonumber\\
&& \left.+ \frac{M_m^2}{(M_m + M_p)^2}\frac{r^2}{R^2}
\left(-\frac{3}{2} + \frac{15}{2} \cos^2(\phi_b -
\phi_p)\right)\right]\label{TOA-expnd-ps}.
\end{eqnarray}
Substituting equations~(\ref{TOA-Vec-Rm}), (\ref{TOA-Vec-Rp}),
(\ref{TOA-expnd-ms}) and (\ref{TOA-expnd-ps}) into (\ref{TOA-d2Rs})
gives, after simplification:
\begin{eqnarray}
\frac{d^2\mathbf{R}_s}{dt^2} &=& \frac{G (M_p +
M_m)}{R^3}\mathbf{R} + \frac{G M_p M_m}{(M_m +
M_p)}\frac{r^2}{R^4} \left[\left(
 -\frac{3}{2} + \frac{15}{2}
\cos^2(\phi_b - \phi_p)\right) \right.
\nonumber\\
&& \Biggl. \times \left(\cos \phi_b\mathbf{i} + \sin
\phi_b\mathbf{j}\right) - 3\cos(\phi_b - \phi_p) \left(\cos \phi_p\mathbf{i} + \sin
\phi_p\mathbf{j}\right)\Biggr]
.\label{TOA-d2Rs2}
\end{eqnarray}
From Figure~\ref{CoordSystem} it can be seen that:
\begin{equation}
\mathbf{n} = \sin I \mathbf{i} + \cos I \mathbf{k}.\label{TOA-ndef}
\end{equation}
Substituting equations~(\ref{TOA-d2Rs2}) and (\ref{TOA-ndef}) into equation~(\ref{TOA-TOApertpmdef}), gives:
\begin{eqnarray}
TOA_{pert,pm} &=&  \frac{- \sin I \ G M_p M_m}{c (M_m +
M_p)}\frac{r^2}{R^4} \left[ \frac{3}{4 n_b^2}\cos \phi_b +
\frac{3}{8(n_b - 2n_p)^2} \cos(\phi_b - 2\phi_p) \right. \nonumber\\
&& \left.+ \frac{15}{8(3n_b - 2n_p)^2} \cos(3\phi_b - 2\phi_p) \right]
.\label{TOA-pert1}
\end{eqnarray}
The $\cos \phi_b$ term in equation~(\ref{TOA-pert1}) has the same
frequency as the signal of a lone planet and it can be shown that it acts to increase the measured value of $M_p +M_m$ derived from $TOA_{pert,p}$ by $(3/4)(r^2/R^2)(M_pM_m/(M_p+M_m))$.  Consequently, this term can be neglected as it will be undetectable as a separate signal. Also, the edge of the
stability region for a prograde satellite of the low-mass component
of a high-mass binary can be approximated by $0.36r_H$ for the case
of circular orbits, where $r_H =
R\left[(M_p)/(3M_s)\right]^{1/3}$ is the secondary's Hill
radius \citep{Holmanetal1999}. When $r$ is equal to this
maximum stable radius $n_p \approx 8n_b$.  As $n_b \ll n_p$ is likely, we have that the denominators of the $\cos(\phi_b - 2\phi_p)$ and $\cos(3\phi_b - 2\phi_p)$ terms will never approach zero.  This, in addition to the assumption of zero eccentricities, means that resonance effects can be neglected. Consequently, equation~(\ref{TOA-pert1}) can be simplified
by neglecting $n_b$ in the denominators, giving:
\begin{equation}
TOA_{pert,pm} =  \frac{- \sin I G M_p M_m}{c (M_m +
M_p)}\frac{r^2}{R^4}\left[\frac{3}{32n_p^2} \cos(\phi_b -
2\phi_p) + \frac{15}{32n_p^2} \cos(3\phi_b - 2\phi_p) \right]
.\label{TOA-pert2}
\end{equation}
Writing $n_p$ in terms of $r$, using Kepler's law, gives:
\begin{equation}
TOA_{pert,pm} =  - \sin I\frac{M_p M_m}{(M_m +
M_p)^2}\frac{R}{c}\left(\frac{r}{R}\right)^5  \left[\frac{3}{32}
\cos(\phi_b - 2\phi_p)  + \frac{15}{32} \cos(3\phi_b - 2\phi_p)
\right] .\label{TOA-pert}
\end{equation}

From equation~(\ref{TOA-pert}), we have that the size of the
perturbation varies as $[M_mM_p/(M_m + M_p)^2][r/R]^5$ times the system crossing time, $R/c$. So, the best hope of a
detectable signal occurs when the planet-moon pair are close to the
parent pulsar, widely separated from each other, both quite massive, and very accurate
timing data is available.

Our result is consistent with a similar study done by \citet{Schneideretal2006}, who calculated the radial velocity perturbation on one component of a binary star system for the case in which the other component was an unresolved pair.  Converting their radial velocity perturbation to a timing perturbation, setting the mass of the planet and moon equal to each other, as in the case investigated by \citet{Schneideretal2006}, and noting that $r$ in this work is equivalent to their $2a_A$, our formula and that of \citet{Schneideretal2006} agree.

\section{Is is Possible to Detect Moons of Planets Orbiting Millisecond Pulsars?}

To investigate whether or not it is possible to detect moons of
pulsar planets, we simplify equation~(\ref{TOA-pert}) by summing the
amplitudes of the sinusoids, giving the maximum possible amplitude:
\begin{equation}\label{ApproxTOAPert}
max\left(TOA_{pert,pm}\right) = \frac{9\sin I}{16}\frac{M_m M_p}{(M_m +
M_p)^2}\frac{R}{c}\left(\frac{r}{R}\right)^5.
\end{equation}
Thus, for a given $r/R$, the maximum amplitude increases linearly with $R$. So,
for planet-moon pairs that are far from each other and their parent
pulsar, detection may be possible.  For example, a 0.1AU
Jupiter-Jupiter binary located 5.2AU from a host pulsar would
produce a  $TOA_{pert,pm}$ of amplitude 960ns, which compares well
with the 130ns residuals obtained from one of the most stable
millisecond pulsars, PSR J0437-4715 \citep{vanStratenetal2001}.

\begin{figure}[tb]
\begin{center}
\includegraphics[height=3in,width=3.84in]{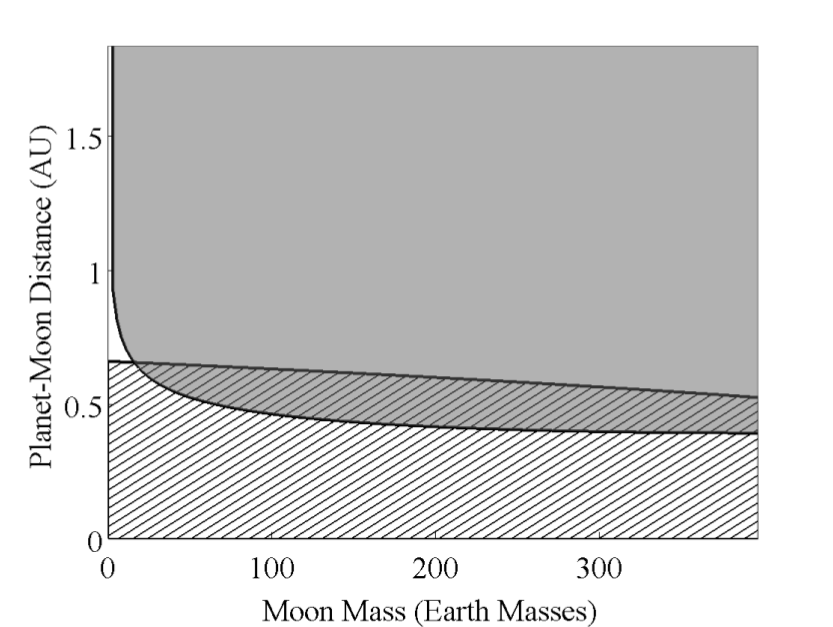}
\caption{The
regions of parameter space containing detectable (shaded) and stable
(cross-hatched) moons of the planet PSR~B1620-26 b are shown as a function of planet-moon separation and moon mass.  The total mass and the
distance of the planet-moon pair from the parent pulsar were assumed
to be 2.5 Jupiter masses and 23AU respectively
\citep{Sigurdssonetal2003}, while it was assumed that $\sin I = 1$.
The mass of the host was set at 1.7 solar masses (the sum of the
mass of the pulsar and its white dwarf companion). The $3 \sigma$
detection threshold was calculated assuming the $\sim 40\mu s$
timing residuals given in \citet{Thorsettetal1999} are uncorrelated
and that similar accuracy TOA measurements of PSR~B1620-26 continue
to the present day.  The stability region was estimated as 0.36
times the Hill sphere of the planet. }
\label{PulsarPlanetMoon}
\end{center}
\end{figure}

To demonstrate this method, the expected maximum signals from a moon
orbiting each of the four known pulsar planets was explored. It was
found that in the case of PSR~B1620-26~b, signals that are in
principle detectable could confirm or rule out certain
configurations of moon mass and orbital parameters (see
Figure~\ref{PulsarPlanetMoon}).

In the particular case of PSR~B1620-26~b, the perturbation signal
will not exactly match the signal shown in equation~(\ref{TOA-pert})
due to the effect of its white dwarf companion.
The effect of this companion was investigated as a side project and
it was found that its effect was to introduce additional
perturbations on top of the $TOA_{pert,p}$ and $TOA_{pert,pm}$
calculated.  Consequently, the detection threshold represents an
upper limit to the minimum detectable signal and thus the analysis
is still valid.

Unfortunately, there are practical limits to the applicability of
this method.  They include discounting other systems that could
produce similar signals, sensitivity limits due to intrinsic pulsar
timing noise, and limits imposed by moon formation and stability.

First, other systems that could produce similar signals need to be
investigated. Possible processes include pulsar precession
\citep*[see e.g.,][]{Akgunetal2006}, periodic variation in the
ISM \citep{Schereretal1997}, gravitational waves
\citep{Detweiler1979}, unmodelled interactions between planets
\citep{Laughlinetal2001} and other small planets. To help
investigate the last two options, we plan on completing a more
in-depth analysis of the perturbation signal of an extra-solar moon,
including the effects of orbital inclination and orbital
eccentricity.

\begin{figure}[tb]
\begin{center}
\includegraphics[height=3in,width=3.84in]{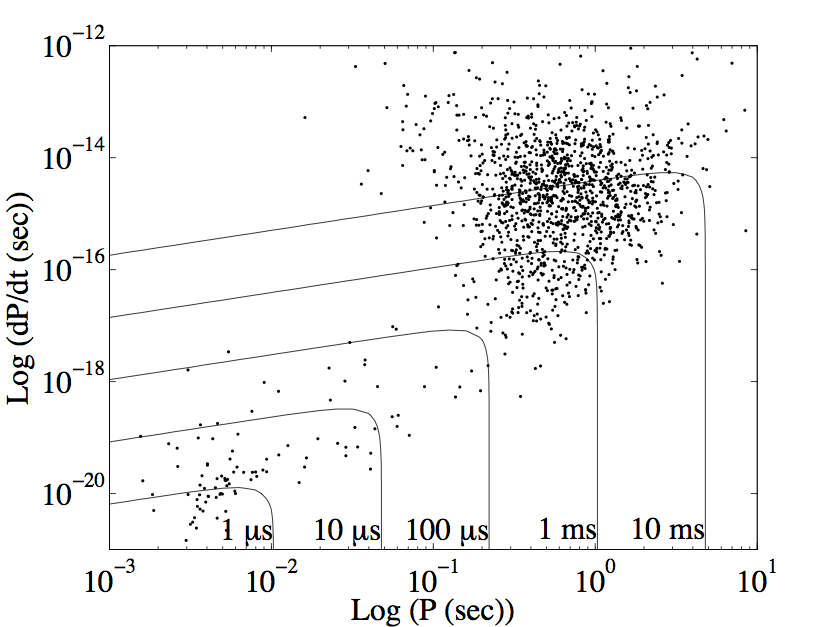}
\caption{Contour plot of
predicted timing noise as a function of pulsar rotation period and
period derivative.  For comparison, all known pulsars as of
publication are overplotted. This plot is based on Figure 9 from
\citet{Cordes1993}. Consequently, the functions and assumptions used
to generate the contours are the same as given in
\citet{Cordes1993}, noting that that each TOA integration is 1000
seconds long. As the correlated timing noise measured for individual
pulsars can vary from the predicted values by two orders of
magnitude \citep[see e.g.,][]{Arzoumanianetal1994}, this plot is
meant to demonstrate global pulsar properties, not predict
individual pulsar noise characteristics. }
\label{PulsarPlot}
\end{center}
\end{figure}

Second, the noise floor of the system needs to be examined. The
suitability of pulsars for signal detection is limited by two main
noise sources, phase jitter and red timing noise \citep[for
example,][]{Cordes1993}. Phase jitter is error due to pulse-to-pulse
variations and leads to statistically independent errors for each TOA
measurement.  Phase jitter decreases with increasing rotation rate
(decreasing $P$) due to the resultant increase in the number of
pulses sampled each integration. Red timing noise refers to noise
for which neigbouring TOA residuals are correlated. Red timing noise has
been historically modeled as a random walk in phase, frequency or
frequency derivative \citep[for
example,][]{Boyntonetal1972,Groth1975,Cordes1980,Kopeikin1997}. Red
noise is strongly dependent on $\dot{P}$.  This relationship can be
understood from the theoretical standpoint that red noise is due to
non-homogeneous angular momentum transport either between components
within the pulsar \citep[e.g.][]{Jones1990} or between it
and its environment \citep[e.g.][]{Cheng1987}. To illustrate
the effect of these two noise sources on TOA accuracy, an estimate
of their combined residuals as a function of $P$ and $\dot{P}$ is
shown in Figure~\ref{PulsarPlot}.  For comparison, the values of $P$
and $\dot{P}$ of every pulsar as of publication are also included.
The ATNF Pulsar
Catalogue\footnote{http://www.atnf.csiro.au/research/pulsar/psrcat/}
\citep{Manchesteretal2005} was used to provide the pulsar data for
this plot.

Third, whether or not moons will be discovered depends on whether or
not they \emph{exist} in certain configurations, which depends on
their formation history and orbital stability. Recent research
suggests that there are physical mass limits for satellites of both
gas giants \citep{Canupetal2006} and terrestrial planets
\citep{Wadaetal2006}. Also, tidal and three-body effects can
strongly affect the longevity of moons
\citep*{Barnesetal2002,Domingosetal2006,Atobeetal2007}.

Finally, while this method was investigated for the specific case of
a pulsar host, this technique could also be applied to planets
orbiting other clock-like hosts such as pulsating giant stars
\citep{Silvottietal2007} and white dwarfs \citep*{Mullallyetal2006}.

\acknowledgments

We are grateful to D. Yong for informing us of planet searches
around pulsating stars.  K. Lewis acknowledges the support of ANU/RSAA where the majority of this work was undertaken.

\end{document}